\documentclass[10pt,doublecolumn]{IEEEtran}
\usepackage{amsmath,amsthm,amsfonts,amssymb,bm,mathrsfs}
\usepackage{graphicx,subfigure}
\usepackage{booktabs,multirow,array,threeparttable}
\usepackage{color}
\usepackage{algorithm,algorithmic}
\usepackage{bbding}
\usepackage{cite,url}

\theoremstyle{remark}

\theoremstyle{plain}

\begin{document}

\title{Deep Reinforcement Learning for Resource Management in Network Slicing}

\author{
Rongpeng Li, Zhifeng Zhao, Qi Sun, Chi-Lin I, Chenyang Yang, Xianfu Chen, Minjian Zhao, and Honggang Zhang

\thanks{R. Li, Z. Zhao, M. Zhao and H. Zhang are with Zhejiang University, Hangzhou 310027, China, (email: \{lirongpeng, zhaozf, mjzhao, honggangzhang\}@zju.edu.cn).}

\thanks{Q. Sun and C.-L. I are with China Mobile Research Institute, Beijing 100053, China (email: \{sunqiyjy, icl\}@chinamobile.com).}

\thanks{C. Yang is with Beihang University, Beijing, 100191, China (email: cyyang@buaa.edu.cn)}

\thanks{X. Chen is with VTT Technical Research Centre of Finland, Oulu FI-90571, Finland (email: xianfu.chen@vtt.fi).}

\thanks{This work was supported in part by National Key R\&D Program of China (No. 2018YFB0803702), National Natural Science Foundation of China (No. 61701439, 61731002), Zhejiang Key Research and Development Plan (No. 2018C03056).}
}

\maketitle

\begin{abstract}
Network slicing is born as an emerging business to operators, by allowing them to sell the customized slices to various tenants at different prices. In order to provide better-performing and cost-efficient services, network slicing involves challenging technical issues and urgently looks forward to intelligent innovations to make the resource management consistent with users' activities per slice. In that regard, deep reinforcement learning (DRL), which focuses on how to interact with the environment by trying alternative actions and reinforcing the tendency actions producing more rewarding consequences, is assumed to be a promising solution. In this paper, after briefly reviewing the fundamental concepts of DRL, we investigate the application of DRL in solving some typical resource management for network slicing scenarios, which include radio resource slicing and priority-based core network slicing, and demonstrate the advantage of DRL over several competing schemes through extensive simulations. Finally, we also discuss the possible challenges to apply DRL in network slicing from a general perspective.
\end{abstract}


\IEEEpeerreviewmaketitle

\section{Introduction}
The fifth-generation cellular networks (5G) is assumed to be the key infrastructure provider for the next decade, by means of profound changes in both radio technologies and network architecture
design \cite{katsalis_network_2017,yousaf_nfv_2017,bega_optimising_2017,li_intelligent_2017}. Besides the pure performance metrics like rate, reliability and allowed connections, the scope of 5G also incorporates the transformation of the mobile network ecosystem and accommodates heterogeneous services using one infrastructure. In order to achieve such a goal, 5G will fully glean the recent advances in the network virtualization and programmability \cite{yousaf_nfv_2017,katsalis_network_2017}, and provide a novel technique named \textit{network slicing} \cite{zhou_network_2016,li_network_2017,katsalis_network_2017,vassilaras_algorithmic_2017}. Network slicing tries to get rid of the current, relatively monolithic architecture like the forth-generation cellular networks (4G) and slice the whole network into different parts, each of which is tailed to meet specific service requirement. Therefore, network slicing is born as an emerging business to operators and allows them to sell the customized network slices to various tenants at different prices. In a word, network slicing could act as a service (NSaaS) \cite{zhou_network_2016}. NSaaS is quite similar to the mature business ``infrastructure as a service (IaaS)'', the benefit of which service providers like Amazon and Microsoft have happily enjoyed for a while. However, in order to provide better-performing and cost-efficient services, network slicing involves more challenging technical issues even for the real-time resource management on existing slices, since (a) for radio access networks, spectrum is a scarce resource and it is meaningful to guarantee the spectrum efficiency (SE) \cite{zhang_network_2017}, while for core networks, virtualized functionalities are limited by computing resources; (b) the service level agreements (SLAs) with slice tenants usually impose stringent requirements on quality of experience (QoE) perceived by users \cite{yu_qos-aware_2017}; and (c) the actual demand of each slice heavily depends on the request patterns of mobile users. Hence, in the 5G era, it is critical to investigate how to intelligently respond to the dynamics of service request from mobile users \cite{vassilaras_algorithmic_2017}, so as to obtain satisfactory QoE in each slice at the cost of acceptable spectrum or computing resources \cite{li_intelligent_2017}. There has been several works towards the resource management for the network slicing, particularly in specific scenarios like edge computing \cite{zanzi_m2ec:_2018} and Internet of things \cite{sciancalepore_slice_2017}. However, it is still very appealing to discuss a approach in generalized scenarios. In that regard, \cite{han_slice_2018} proposes to adopt genetic algorithm as an evolutionary means for inter-slice resource management. However, \cite{han_slice_2018} does not reflect the explicit relationship that one slice might require more resources due to its more stringent SLA.

On the other hand, partially inspired by the psychology of human learning, the learning agent in reinforcement learning (RL) algorithm focuses on how to interact with the environment (represented by \textit{states}) by trying alternative \textit{actions} and reinforcing the tendency actions producing more rewarding consequences \cite{sutton_reinforcement_1998}. Besides, reinforcement learning also embraces the theory of optimal control and adopts some ideas like value functions and dynamic programming. However, reinforcement learning faces some difficulties in dealing with large state space, since it is challenging to traverse every state and obtain a value function or model for every station-action pair in a direct and explicit manner. Hence, benefiting from the advances in graphics processing units (GPUs) and the less concern for the computing power, some researchers propose to sample only a fraction of states and further apply neural networks (NN) to train a sufficiently accurate value function or model. Following this idea, Google DeepMind has pioneered to combine NN with one typical RL algorithm (i.e., $\mathcal{Q}$-Learning), and proposed one deep reinforcement learning (DRL) algorithm with enough performance stabilities \cite{silver_mastering_2016,mnih_human-level_2015}.

The well-known success of AlphaGo \cite{silver_mastering_2016} and following exciting results to apply DRL to solve resource allocation issues in some specific fields like power control \cite{nasir_deep_2018}, green communications \cite{liu_deepnap:_2018}, cloud radio access networks \cite{xu_deep_2017}, mobile edge computing and caching \cite{he_software-defined_2017,he_deep-reinforcement-learning-based_2017,he_integrated_2018}, have aroused some research interest to apply DRL to the field of network slicing. However, given the challenging technical issues in the resource management on existing slices, it is critical to carefully investigate the performance of applying DRL in the following aspects:
\begin{itemize}
	\item The basic concern is whether or not the application of DRL is feasible. More specifically, does DRL produce satisfactory QoE results while consuming acceptable network resources (e.g., spectrum)?
	\item The research community has proposed some schemes for the resource management in network slicing scenarios. For example, the resource management could be conducted by either following a meticulously designed prediction algorithm, or equally dividing the available resource into each slice. The former implies one reasonable option, while the latter saves a lot of computational cost. Hence, a comparison between DRL and these interesting schemes is also necessary.
\end{itemize}
In this paper, we strive to address these issues.

The remainder of the paper is organized as follows. Section \ref{sec:ALGORITHM DESCRIPTION} starts with the fundamentals of RL and talks about the motivation to evolve towards DRL from RL. As the main part of the paper, Section \ref{sec:resource} addresses two resource management issues in network slicing and highlights the advantages of DRL by extensive simulation analyses. Section \ref{sec:conclusion} concludes the paper and points out some research directions to apply DRL in a general manner.

\section{From Reinforcement Learning to Deep Reinforcement Learning}
\label{sec:ALGORITHM DESCRIPTION}
In this section, we give a brief introduction over RL or more specifically $\mathcal{Q}$-Learning, and then talk about the motivation to evolve from $\mathcal{Q}$-Learning to Deep $\mathcal{Q}$-Learning (DQL).

\begin{figure*}
	\centering
	\includegraphics[width = 0.80\textwidth]{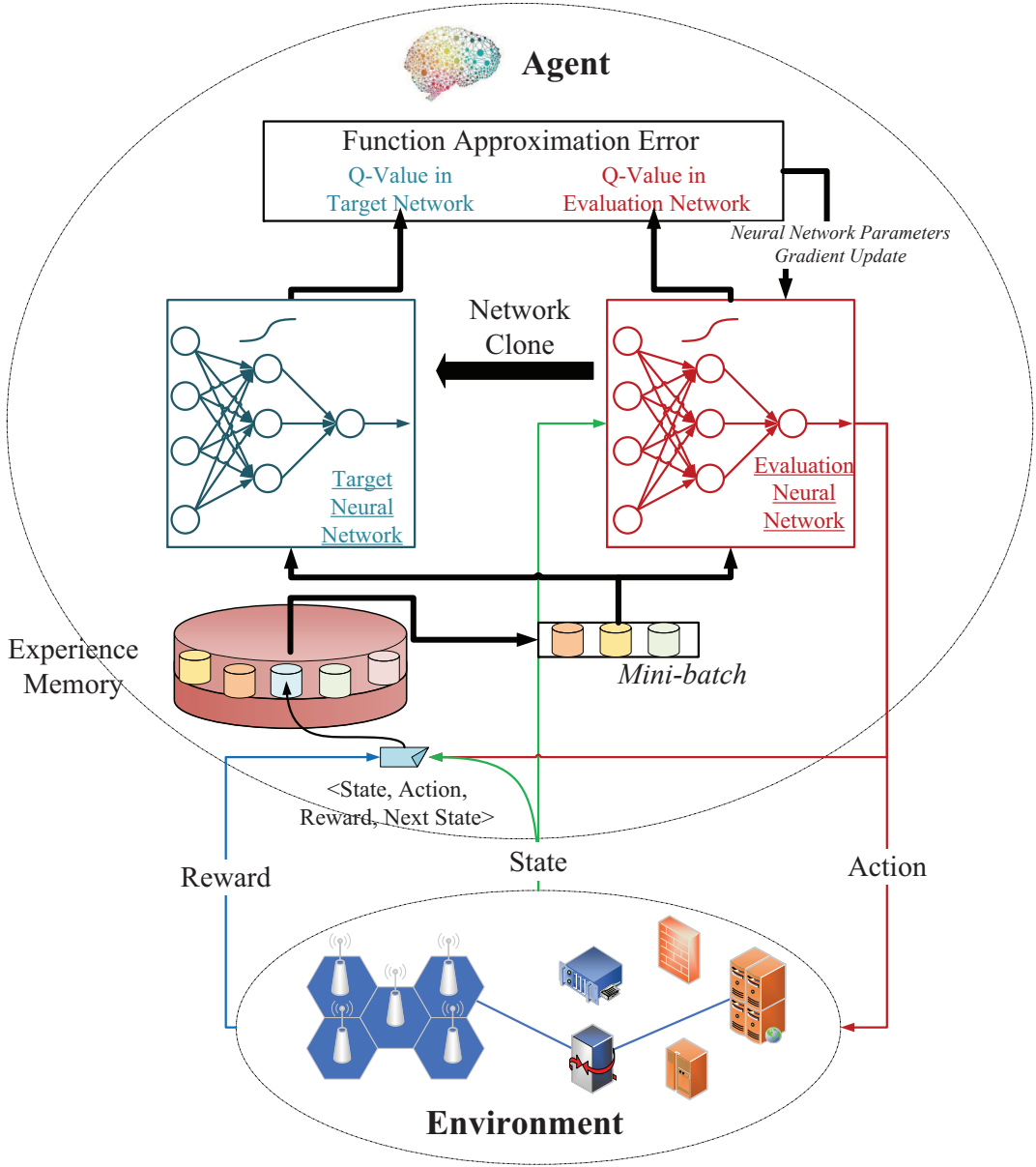}
	\caption{An illustration of deep $\mathcal{Q}$-learning.}
	\label{fig:learning}
\end{figure*}
\subsection{Reinforcement Learning}
RL learns how to interact with the environment to achieve maximum cumulative return (or average return), and has been successfully applied in the fields like robot control, self driving, and chess playing for years. Mathematically, RL follows the typical concept of Markov decision process (MDP), while the MDP is a generalized framework for modeling decision-making problems in cases where the result is partially random and affected by the applied decision. An MDP can be formulated by a $5$-tuple as $M = \langle S,A,P(s'|s,a),R,\gamma \rangle$, where $S$ and $A$ denote a finite state space and action set, respectively. $P(s'|s,a)$ indicates the probability that the action $a \in A$ under state $s \in S$ at slot $t$ leads to state $s' \in S$ at slot $t+1$. $R(s,a)$ is an immediate reward after performing the action $a$ under state $s$, while $\gamma \in [0,1]$ is a discount factor to reflect the diminishing importance of current reward on future ones. Usually, the goal of MDP is to find a policy $a = \pi(s)$ that determines the selected action $a$ under state $s$, so as to maximize the value function, which is typically defined as the expected discounted cumulative reward by the Bellman equation:
\begin{equation}
	\label{eq:definition_state_value_function}
	\begin{aligned}
	V^{\pi}(\hat{s}) &=E_{\pi}\left[\sum\limits_{k=0}^{\infty} \gamma^k R(s^{(k)},\pi(s^{(k)}))|s^{(0)}=\hat{s})\right]\\
	&=E_{\pi}\left[R(\hat{s},\pi(\hat{s})))+\gamma \sum_{\bm{s}' \in \mathcal{S}} P(s'|\hat{s},\pi(\hat{s}))V^{\pi}(s')\right].
	\end{aligned}
\end{equation}

Dynamic programming could be exploited to solve the Bellman equation when the state transition probability $P(s'|s,a)$ is known apriori with no random factors. But inspired by both control theory and behaviorist psychology, RL aims to obtain the optimal policy $\pi^\ast$ under circumstances with unknown and partially random dynamics. Since RL does not have explicit knowledge over whether it has come close to its goal, it needs the balance between exploring new potential actions and exploiting the already learnt experience. So far, there has been some classical RL algorithms like $\mathcal{Q}$-learning, actor-critic method, SARSA, TD($\lambda$), etc \cite{sutton_reinforcement_1998}. Given by the detailed methodologies and practical application scenarios, we can classify these RL algorithms according to different criteria:
\begin{itemize}
	\item \textit{Model-based versus Model-free}: Model-based algorithms imply the agent tries to learn the model of how the environment works from its observations and then plan a solution using that model. Once the agent gains adequately accurate model, it can use a planing algorithm with its learned model to find a policy. Model-free algorithms means the agent does not directly learn how to model the environment. Instead, like the classical example of $\mathcal{Q}$-learning, the agent estimates the Q-values (or roughly the value function) of each state-action pair and derives the optimal policy by choosing the action yielding the largest Q-value in the current state. Different from the model-based algorithm, the well-learnt model-free algorithm like $\mathcal{Q}$-learning cannot predict the next state and value before taking the action.
	\item \textit{Monte-Carlo Update versus Temporal-Difference Update}: Generally, the value function update could be conducted in two ways, that is, the Monte-Carlo update and the temporal-difference (TD) update. A Monte-Carlo update means the agent updates its estimation for a state-action pair by calculating the mean return from a collection of episodes. A TD update approximates the estimation by comparing estimates at two consecutive episodes. For example, $\mathcal{Q}$-learning updates its Q-value by the TD update as $Q(s,a) \leftarrow Q(s,a) + \alpha (R(s,a) + \gamma \max_{a'} Q(s',a') - Q(s,a) )$, where $\alpha$ is the learning rate. Specifically, the term $R(s,a) + \gamma \max_{a'} Q(s',a') - Q(s,a) $  is also named as the TD error, since it captures the difference between the current (sampled) estimate $R(s,a) + \gamma \max_{a'} Q(s',a')$ and previous one $Q(s,a)$.
	\item \textit{On-policy versus Off-policy}: The value function update is also coupled with the executed update policy. Before updating the value function, the agent also needs to sample and learn the environment by performing some non-optimal policy. If the update policy is irrelevant to the sampling policy, the agent is called to perform an off-policy update. Taking the example of $\mathcal{Q}$-learning, this off-policy agent updates the Q-value by choosing the action corresponding to the best Q-value, while it could learn the environment by adopting sampling policies like $\epsilon$-greedy or Boltzmann distribution to balance the ``exploration and exploitation" problem \cite{sutton_reinforcement_1998}. The $\mathcal{Q}$-learning proves to converge regardless of the chosen sampling policy. On the contrary, the SARSA agent is on-policy, since it updates the value function by $Q(s,a) \leftarrow Q(s,a) + \alpha (R(s,a) + \gamma Q(s',a') - Q(s,a) )$ where $a'$ and $a$ need to be chosen according to the same policy. 
\end{itemize}

\subsection{From $\mathcal{Q}$-Learning to Deep $\mathcal{Q}$-Learning}
\label{sec:Deep neural networks}
We first summarize the details of $\mathcal{Q}$-Learning. Generally speaking, $\mathcal{Q}$-Learning belongs to a model-free, TD update, off-policy RL algorithm, and consists of three major steps:
\begin{enumerate}
	\item The agent chooses an action $a$ under state $s$ according to some policy like $\epsilon$-greedy. Here, the $\epsilon$-greedy policy means the agents chooses the action with the largest Q-value $Q(s,a)$ with a probability of $\epsilon$, and equally chooses the other actions with a probability of $\frac{1-\epsilon}{|A|}$, where $|A|$ denotes the size of the action space.
	\item The agent learns the reward $R(s,a)$ from the environment, and the state transitions to the next state $s'$.
	\item The agent updates the Q-value function in a TD manner as $Q(s,a) \leftarrow Q(s,a) + \alpha\left( R(s,a) + \gamma \max_{a'} Q(s',a') - Q(s,a) \right)$.
\end{enumerate}

Classical RL algorithms usually rely on two different ways (i.e., explicit table or function approximation) to store the estimated value functions. For the table storage, RL algorithm uses an array or hash table to store the learnt results for each state-action pair. For large state space, it not only requires intensive storage, but also is unable to quickly transverse the complete the state-action pair. Due to the curse of dimensionality, function approximation sounds more appealing. 

The most straightforward way for function approximation is a linear approach. Taking the example of $\mathcal{Q}$-learning, the Q-value function could be approximated by a linear combination of $n$ orthogonal bases $ \bm{\psi}(s,a) = \{\psi_1(s,a), \cdots \psi_n(s,a)\}$, that is,
$	Q(s,a) = \theta_0 \cdot 1 +  \theta_1 \cdot \psi_1(s,a) + \cdots  +  \theta_n \cdot \psi_n(s,a)  = \bm{\theta}^T \bm{\psi}(s,a) $, where $\theta_0$ is a biased term with $1$ absorbed into the $\bm{\psi}$ for simplicity of representation and $\bm{\theta}$ is a vector with the dimension of $n$. The function approximation in the $\mathcal{Q}$-learning means that $Q(s,a) = \bm{\theta}^T \bm{\psi}(s,a) $ should be as close as the learnt ``target" value $Q^{+}(s,a) = \sum_{s} P(s'| s,a) \big[ R(s,a) + \gamma \max_{a'} Q^{+}(s',a') \big]$ over all the state-action pairs. Since it is infeasible to transverse all the state-action pairs, the ``target" value could be approximated based on the minibatch samples and  $Q^{+}(s,a) \approx  R(s,a) + \gamma \max_{a'} Q^{+}(s',a') $. In order to make $Q(s,a) = \bm{\theta}^T \bm{\psi}(s,a) $ approach the ``target" value $Q^{+}(s,a)$, the objective function could be defined as
\begin{align}
	& L(\bm{\theta}) \nonumber \\
	= &  \frac{1}{2} \big( Q^{+}(s,a) - Q(s,a)\big)^2 \label{eq:loss_function} \\
	= & \frac{1}{2} \big( Q^{+}(s,a) -  \bm{\theta}^T \bm{\psi}(s,a) \big)^2. \nonumber
\end{align}

The parameter $\bm{\theta}$ minimizing $L(\bm{\theta}) $ could be achieved by a gradient-based approach as 
\begin{align}
	\label{eq:gradient}
	&\bm{\theta}^{(i+1)} \nonumber \\
	\leftarrow & \bm{\theta}^{(i)} - \alpha \nabla L(\bm{\theta}^{(i)}) \\
	= &  \bm{\theta}^{(i)} - \alpha  \big( Q^{+}(s,a) -  \bm{\theta}^T \bm{\psi}(s,a) \big) \bm{\psi}(s,a). \nonumber
\end{align}
For a large state-action space, the function approximation reduces the number of unknown parameters to a vector with dimension $n$ and the related gradient method further solves the parameter approximation in an computationally efficient manner.

Apparently, the linear function approximation could not accurately model the estimated value function.
Hence, researchers have proposed to replace the approximation $Q(s,a;\bm{\theta})$ by some non-linear means. In that regard, NN is skilled in approximating non-linear functions \cite{hornik_multilayer_1989}. Therefore, in AlphaGo \cite{silver_mastering_2016,mnih_human-level_2015}, NN has been exploited and the loss function can be re-defined as $L(\bm{\theta}) = \frac{1}{2} \big( Q^{+}(s,a) - Q(s,a; \bm{\theta})\big)^2$. Besides, deep neural network has made novel progress in the following aspects:
\begin{itemize}
	\item \textit{Experience Replay}\cite{mnih_human-level_2015}: The agent stores the past experience (i.e., the tuple $e_t = \langle s_t,a_t, s'_t, R(s_t,a_t)\rangle$) at episode $t$ into a dataset $D_t = (e_1,\cdots,e_t)$ and uniformly selects some (mini-batch) items from the dataset to update the Q-value neural network $Q(s,a;\bm{\theta})$. 
	\item \textit{Network Cloning}: The agent uses a separate network $\hat{Q}$ to guide how to select an action $a$ in state $s$, and the network $\hat{Q}$ is replaced by $Q$ every $C$ episodes. Simulation results demonstrate that this network cloning enhances the learning stability \cite{mnih_human-level_2015}.
\end{itemize}
Both experience replay and network cloning motivate to choose the off-policy $\mathcal{Q}$-learning, since the sampling policy is only contingent on previously trained Q-value NN and the updating policy, which relies on the information from the new episodes, is irrespective of the sampling policy. On the other hand, the DQL agent could collect the information (i.e., state-action-reward pair) and train its policy in background. Also, the learned policy is stored in the neural networks and can be conveniently transferred among similar scenarios. In other words, the DQL could efficiently perform and timely make the resource allocation decision according to its already learned policy.

Finally, we illustrate the deep $\mathcal{Q}$-learning in Fig. \ref{fig:learning} and summarize the general steps in Algorithm \ref{al:originalDQL}.
\begin{algorithm}
	\caption{The general steps of deep reinforcement learning.}
	\label{al:originalDQL}
	\begin{algorithmic}[1]
	\REQUIRE An evaluation network $Q$ with weights $\bm{\theta}$; a target network $\hat{Q}$ with weights $\hat{\bm{\theta}} = \bm{\theta}$.\\   
	\ENSURE A replay memory dataset $D$ with size of $N$; the episode index $t=0$.\\
	\REPEAT
		\STATE At episode $t$, the DQL agent observes the state $s_t$. 
		\STATE The agent chooses action $a_t$ with a probability $\epsilon$ or selects $a_t$ satisfying $a_t = \arg \max_a Q (s_t, a;\bm{\theta})$.\\
		\STATE After executing the action $a_t$, the agent observes the reward $R(s_t,a_t)$ and a new state $s_{t+1} = s'_t$ for the system.
		\STATE The agent stores the episode experience $e_t =  \langle s_t,a_t, s'_t, R(s_t,a_t)\rangle $ into $D$.
		\STATE The agent samples a minibatch of experiences from $D$ and sets $Q^{+}(s_t,a_t) = R(s_t,a_t) + \gamma \max_{a'} Q^{+}(s'_t,a')$. In cases where episode terminates at $t$, $Q^{+}(s_t,a_t) = R(s_t,a_t)$.
		\STATE The agent updates the weights $\bm{\theta}$ for the evaluation network by a gradient-based approach in \eqref{eq:gradient}.
		\STATE The agent clones the evaluation network $Q$ to the target network $\hat{Q}$ every $C$ episodes by assigning the weights $\hat{\bm{\theta}}$ as $\hat{\bm{\theta}} = \bm{\theta}$.
		\STATE The episode index is updated by $t \leftarrow t+1$.
	\UNTIL{A predefined stopping condition (e.g., the gap between $\bm{\theta}$ and $\hat{\bm{\theta}}$, the episode length, etc) is satisfied.}
	\end{algorithmic}
\end{algorithm}

\section{Resource Management for Network Slicing}   
\label{sec:resource}      
\begin{figure*}
	\centering
	\includegraphics[width=0.895\textwidth]{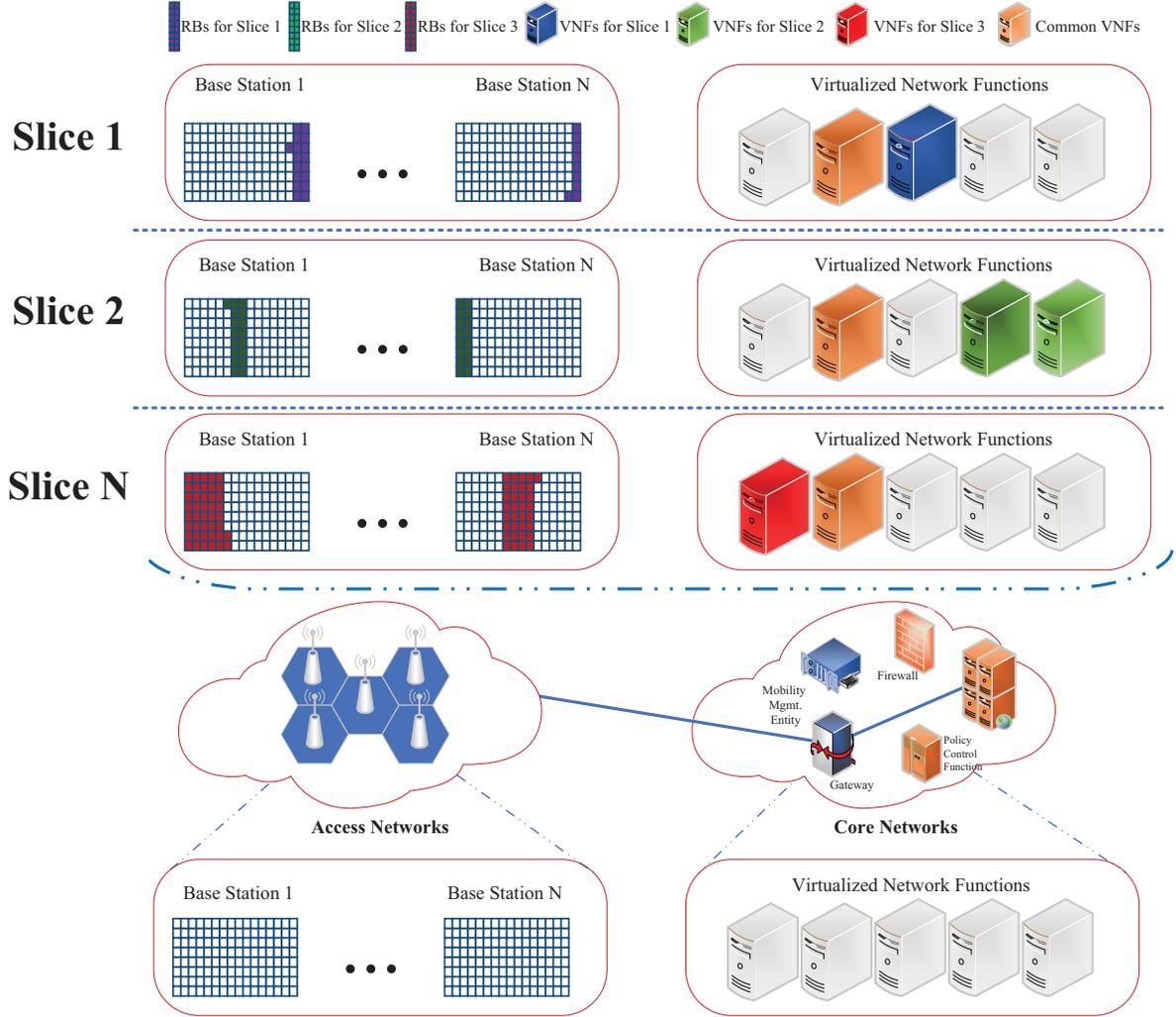}
	\caption{An illustration of resource management for network slicing.}
	\label{fig:slicing}
\end{figure*}

Resource management is a permanent topic during the evolution of wireless communication. Intuitively, resource management for network slicing can be considered from several different perspectives.

\begin{itemize}  
	\item \textit{Radio Resource and Virtualized Network Functions}: As depicted in Fig. \ref{fig:slicing}, resource management for network slicing involves both radio access part and core network part with slightly different optimization goals. Due to the limited spectrum resource, the resource management for the radio access puts considerable efforts in allocating resource blocks (RBs) to one slice, so as to maintain acceptable SE while trying to bring appealing rate and small delay. The widely adopted optical transmission in core networks has shifted the optimization of core network to design common or dedicated virtualized network functions (VNFs), so as to appropriately forward the packets from one specific slice with minimal scheduling delay. By balancing the relative importance of resource utilization (e.g, SE) and QoE satisfaction ratio, the resource management problem could be formulated as $R = \zeta \cdot \text{SE} + \beta \cdot \text{QoE}$, where $\zeta$ and $\beta$ denotes the importance of SE and QoE.
	\item \textit{Equal or Prioritized Scheduling}: As part of the control plane, IETF \cite{de_foy_network_2017} has defined the common control network function (CCNF) to all or several slices. The CCNF includes the access and mobility management function (AMF) as well as the network slice selection function (NSSF), which is in charge of selecting core network slice instances. Hence, besides equally treating flows from different slices, the CCNF might differentiate flows. For example, flows from ultra-reliable low-latency communications (URLLC) service can be scheduled and provisioned in higher priority, so as to experience as little latency as possible. In this case, in order to balance the resource utilization (RU) and the waiting time (WT) of flows, the objective goal could be similarly written as a weighted summation of RU and WT.
\end{itemize} 

Based on the aforementioned discussions, we can safely reach a conclusion that, the objective of resource management for network slicing should take account of several variables and a weighted summation of these variables can be considered as the reward for the learning agent.

\begin{table*}  
\centering
\caption{A Brief Summary of Key Settings in DRL for Network Slicing Simulations}  
\label{tab:mapping}  
\begin{tabular}{m{1.5cm} | m{5.5cm} | m{5.5cm} } 
	\toprule[1.5pt] 
	\multicolumn{3}{c}{(a) The Mapping from Resource Management for Network Slicing to DRL}\\
	\midrule[0.8pt] 
	& Radio Resource Slicing & Priority-based Core Network Slicing  \\  
	\midrule[0.8pt] 
	State & The number of arrived packets in each slice within a specific time window & The priority and time-stamp of last arrived five flows in each service function chain (SFC) \\  
	\hline
	Action & Allocated bandwidth to each slice & Allocated SFC for the flow at current time-stamp \\  
	\hline
	Reward & Weighted sum of SE and QoE in 3 sliced bands  & Weighted sum of average time in 3 SFCs\\
	\bottomrule[1.5pt] 
\end{tabular}  
\begin{tabular}{m{3cm} | m{3cm} | m{3cm} | m{3cm} } 
	\multicolumn{4}{c}{(b) Parameter settings for radio resource slicing}\\
	\midrule[0.8pt] 
	& VoLTE & Video & URLLC  \\  
	\midrule[0.8pt] 
	Bandwidth & \multicolumn{3}{l}{10 MHz}\\
	\hline
	Scheduling & \multicolumn{3}{l}{Round robin per slot (0.5 ms)}\\
	\hline
	Slice Band Adjustment (Q-Value Update) & \multicolumn{3}{l}{1 second (2000 scheduling slots) }\\
	\hline
	Channel & \multicolumn{3}{l}{Rayleigh fading}\\
	\hline 
	User No. (100 in all) & 46 & 46 & 8\\
	\hline
	Distribution of Inter-Arrival Time & Uniform [Min = 0, Max = 160ms] & Truncated Pareto [Exponential Para = 1.2, Mean = 6 ms, Max = 12.5 ms] & Exponential [Mean = 180 ms] \\
	\hline
	Distribution of Packet Size & Constant (40 Byte) &  Truncated Pareto [Exponential Para = 1.2, Mean = 100 Byte, Max = 250 Byte]  & Truncated Lognormal
	[Mean = 2 MB, Standard Deviation = 0.722 MB, Maximum =5 MB] \\
	\hline 
	SLA: Rate & 51 kbps & 5 Mbps & 10 Mbps \\
	\hline
	SLA: Latency & 10 ms & 10 ms & 5 ms\\
	\bottomrule[1.5pt] 
\end{tabular}  
\end{table*} 

\subsection{Radio Resource Slicing}
\label{sec:radio}
In this part, we address how to apply DRL for radio resource slicing. Mathematically, given a list of existing slices $1, \cdots, N$ sharing the aggregated bandwidth $W$ and having fluctuating demands $\bm{d}=(d_1, \cdots, d_N)$, DQL tries to give a bandwidth sharing solution $\bm{w}=(w_1, \cdots, w_N)$, so as to maximize the long-term reward expectation $\mathbb{E}\{R(\bm{w},\bm{d})\}$ where the notation $\mathbb{E}(\cdot)$ denotes to take the expectation of the argument, that is,

\begin{align}
	&\arg_{\bm{w}} \max \mathbb{E} \{R(\bm{w},\bm{d})\} \nonumber\\
	= & \arg_{\bm{w}} \max \mathbb{E}\big\{\zeta \cdot \text{SE} (\bm{w},\bm{d})+ \beta \cdot \text{QoE} (\bm{w},\bm{d}) \big\} \nonumber\\
	\text{s.t.: } & \bm{w}=(w_1, \cdots, w_N) \label{eq:formulation}\\
	& w_1+ \cdots + w_N = W \nonumber\\	
	& \bm{d}=(d_1, \cdots, d_N )\nonumber\\
	& d_i \sim  \text{Certain Traffic Model}, \forall i \in [1, \cdots, N] \nonumber
\end{align}
The key challenge to solve \eqref{eq:formulation} lies in the volatile demand variations without having known a priori due to the traffic model. Hence, DQL is exactly the matching solution to solve the problem.

We evaluate the performance to adopt DQL to solve \eqref{eq:formulation} by simulating a scenario containing one single BS with three types of services (i.e., VoIP, video, URLLC). There exist 100 registered subscribers randomly located within a 40 meter-radius circle surrounding the BS. These subscribers generate service models summarized in Table \ref{tab:mapping}(b). VoIP and video services exactly take the parameter settings of VoLTE and video streaming models, while URLLC service takes the parameter settings of FTP 2 model \cite{ngmn_ngmn_nodate}. It can be observed from Table \ref{tab:mapping}(b), URLLC has less frequent packets compared with the others, while VoLTE requires the smallest bandwidth for its packets.

We consider DQL by using the mapping in Table \ref{tab:mapping}(a) to optimize the weighted summation of system SE and slice QoE. Specifically, we perform round-robin scheduling method within each slice at the granularity of 0.5 ms. In other words, we sequentially allocate the bandwidth of each slice to the active users within each slice every 0.5 ms. Besides, we adjust the bandwidth allocation to each slice per second. Therefore, the DQL agent updates its Q-value neural network every second. We compare the simulation results with the following three methods, so as to explain the importance of DQL.
\begin{itemize}
\item \textit{Demand-prediction based method}: The method tries to estimate the possible demand by using long short-term memory (LSTM) to predict the number of active users requesting VoIP, video and URLLC respectively. Afterwards, the bandwidth is allocated by two ways: (1) \textit{DP-No} allocates the whole bandwidth to each slice proportional to the number of predicted packets. In particular, assuming that  the total bandwidth is $B$ and the predicted number of packets for VoIP, video and URLLC is $N_{\rm{VoIP}}$, $N_{\rm{Video}}$ and $N_{\rm{URLLC}}$, the allocated bandwidth to these three slices (i.e., VoIP, video and URLLC) is $\frac{B\cdot N_{\rm{VoIP}}}{N_{\rm{VoIP}}+ N_{\rm{Video}} + N_{\rm{URLLC}}}$, $\frac{B\cdot N_{\rm{Video}}}{N_{\rm{VoIP}}+ N_{\rm{Video}} + N_{\rm{URLLC}}}$, $\frac{B\cdot N_{\rm{URLLC}}}{N_{\rm{VoIP}}+ N_{\rm{Video}} + N_{\rm{URLLC}}}$, respectively. (2) \textit{DP-BW} performs the allocation by multiplying the number of predicted packets by the least required rate in Table \ref{tab:mapping}(b) and then computing the proportion. In this regard, assuming that the required rate for the three slices is $R_{\rm{VoIP}}$, $R_{\rm{Video}}$ and $R_{\rm{URLLC}}$, the allocated bandwidth to VoIP, video and URLLC is $\frac{B N_{\rm{VoIP}} R_{\rm{VoIP}}}{N_{\rm{VoIP}}R_{\rm{VoIP}}+ N_{\rm{Video}} R_{\rm{Video}}+ N_{\rm{URLLC}}R_{\rm{URLLC}}}$, $\frac{B N_{\rm{Video}} R_{\rm{Video}}}{N_{\rm{VoIP}}R_{\rm{VoIP}}+ N_{\rm{Video}} R_{\rm{Video}}+ N_{\rm{URLLC}}R_{\rm{URLLC}}}$, $\frac{B N_{\rm{URLLC}} R_{\rm{URLLC}}}{N_{\rm{VoIP}}R_{\rm{VoIP}}+ N_{\rm{Video}} R_{\rm{Video}}+ N_{\rm{URLLC}}R_{\rm{URLLC}}}$, respectively. Round-robin is conducted within each slice.
\item \textit{Hard slicing}: Hard slicing means that each service slice is always allocated $\frac{1}{3}$ of the whole bandwidth, since there exists 3 types of service in total. Again, round-robin is conducted within each slice.
\item \textit{No slicing}: Irrespective of the related SLA, all users are scheduled equally. Round-robin is conducted within all users.
\end{itemize} 
We primarily consider the downlink case and adopt system SE and QoE satisfaction ratio as the evaluation metrics. In particular, the system SE is computed as the number of bits transmitted per second per unit bandwidth, where the rate from the BS to users is derived based on Shannon capacity formula. Therefore, if part of the bandwidth has been allocated to one slice but the slice has no service activities at one slot, such part of bandwidth has been wasted, thus degrading the system SE. QoE satisfaction ratio is obtained by dividing the number of completely transmitted packets satisfying rate and latency requirement by the total number of arrived packets.
\begin{figure*}[!t]
	\centering
	\includegraphics[width=.975\textwidth]{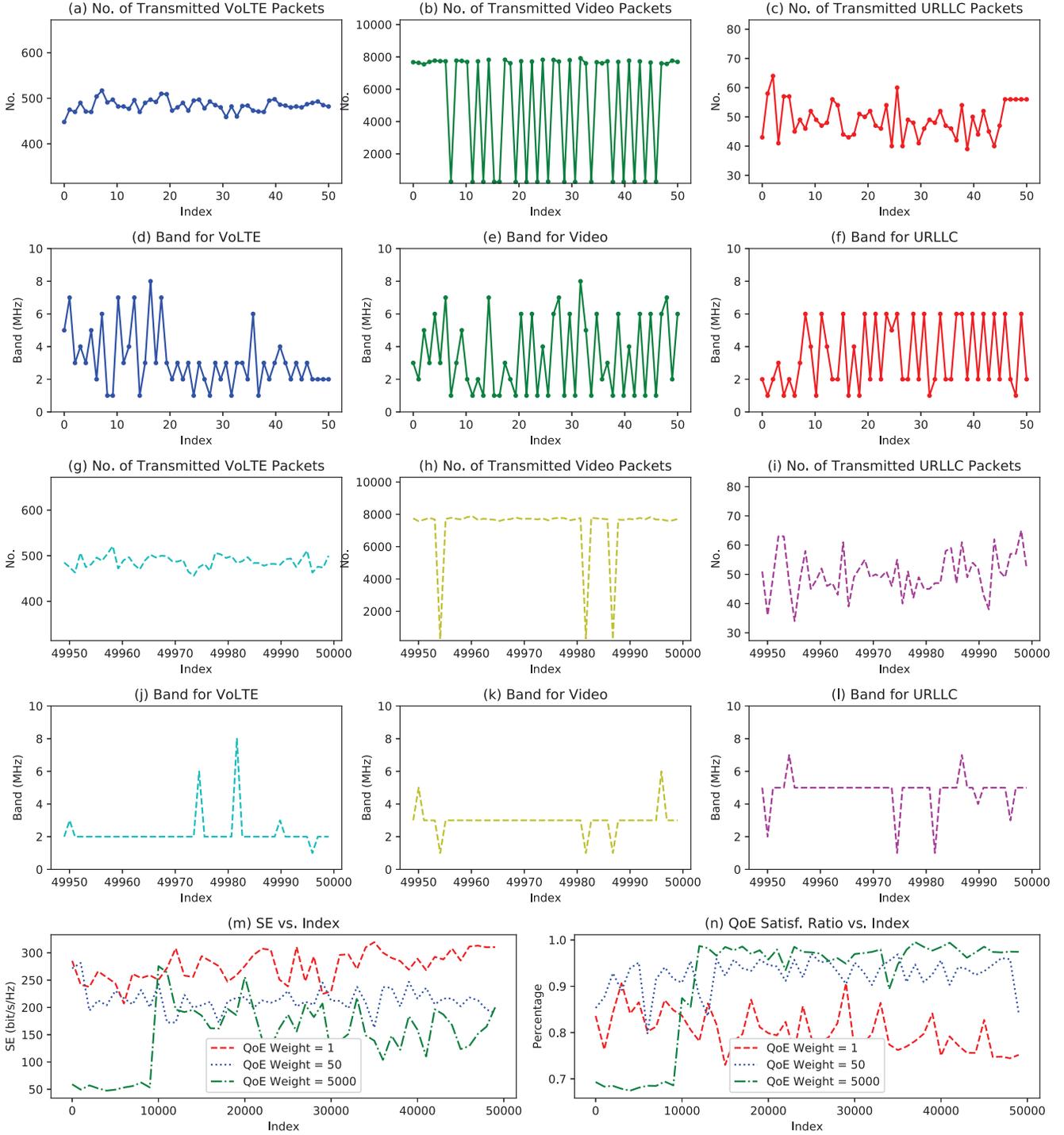}
	\caption{The performance of DQL for radio resource slicing w.r.t. the learning steps (QoE Weight = 5000).}
	\label{fig:learning_r} 
\end{figure*}

Fig. \ref{fig:learning_r} presents the learning process of DQL\footnote{Notably, $\gamma$ is set as $0.9$.} in radio resource management. In particular, Fig. \ref{fig:learning_r}(a)$\sim$\ref{fig:learning_r}(f) give the initial performance of DQL when the QoE weight is $5000$ and the SE weight is $0.1$. Fig. \ref{fig:learning_r}(g)$\sim$\ref{fig:learning_r}(l) provide the performance during the last 50 of 50000 learning updates. From these sub-figures, it can be observed that DQL could not well learn the user activities at the very beginning and the allocated bandwidth fluctuates heavily. But after nearly 50000 updates, DQL has gained better knowledge over user activities and yielded a state bandwidth-allocation strategy. Besides, Fig. \ref{fig:learning_r}(m) and Fig. \ref{fig:learning_r}(n) show the variations of SE and QoE along with each learning epoch. From both subfigures, a larger QoE weight produces policies with superior QoE performance while bringing certain loss in the system SE performance.

\begin{figure*}[!t]
	\centering
	\includegraphics[width=.975\textwidth]{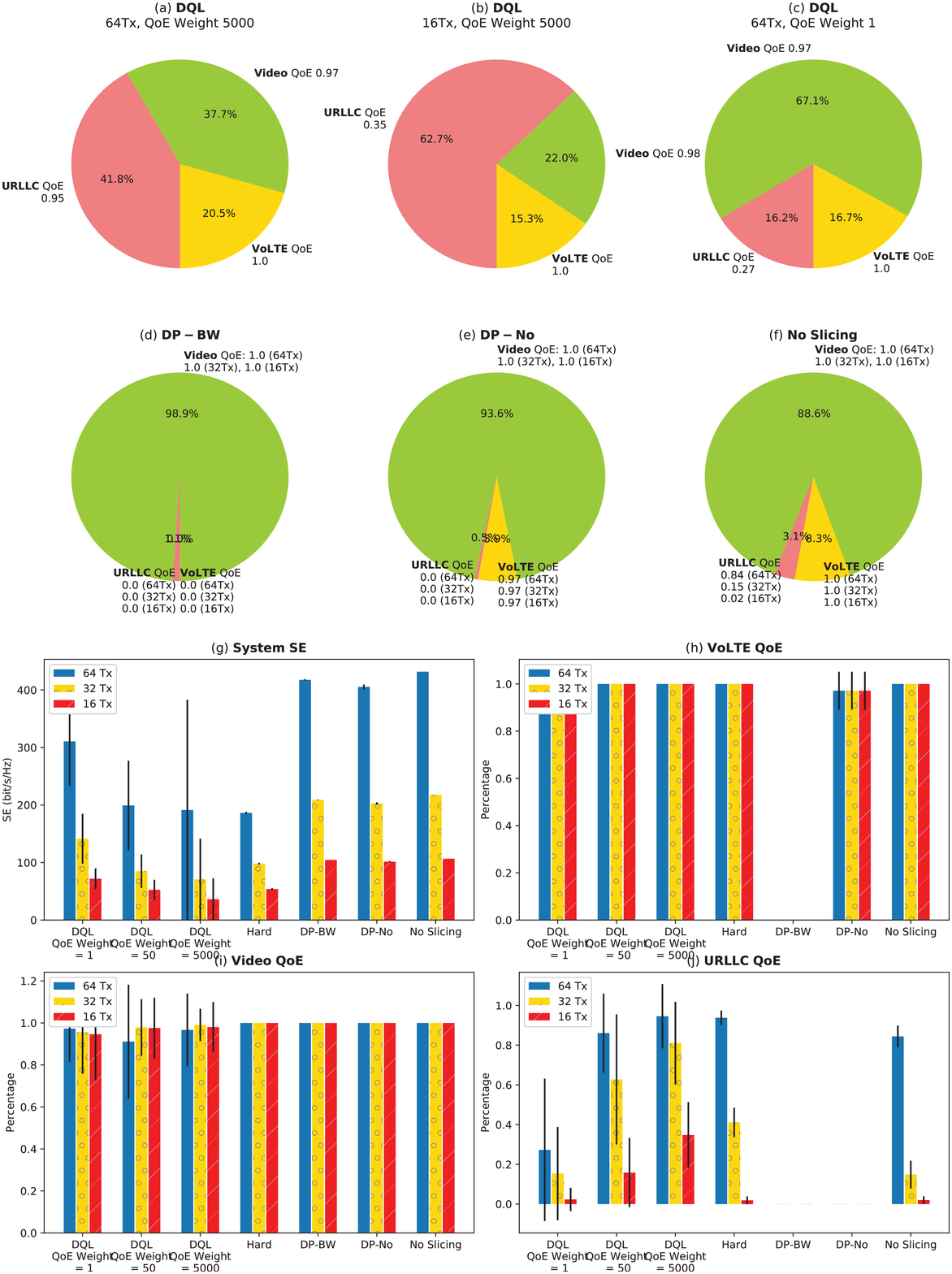}
	\caption{The performance comparison among different schemes for radio resource slicing.}
	\label{fig:compare} 
\end{figure*}

Fig. \ref{fig:compare} provides a detailed performance comparison among the candidate techniques, where the results for DQL are obtained after 50000 learning updates. Fig. \ref{fig:compare}(a)$\sim$\ref{fig:compare}(f) gives the percentage of total bandwidth allocated to each slice using the pie charts and highlights the QoE satisfaction ratio by surrounding text. From Fig. \ref{fig:compare}(a)$\sim$\ref{fig:compare}(b), a reduction in transmission antennas from 64 to 16, which implies a decrease in network capability and an increase in potential collisions across slices, leads to a re-allocation of network bandwidth inclined to the bandwidth-consuming yet activity-limited URLLC slice. Also, it can be observed from Fig. \ref{fig:compare}(f), when the downlink transmission uses 64 antennas, ``no slicing" performs the best, since the transmission capability is sufficient and the scheduling period is 0.5 ms while the bandwidth allocated to each slice is adjusted per second and thus slower to catch the demand variations. When the number of downlink antenna turns to 32, the DQL-driven scheme produces 81\% QoE satisfaction ratio for URLLC, while ``no slicing" and ``hard slicing" schemes only provision 15\% and 41\% satisfied URLLC packets, respectively. Notably, applying DQL mainly leads to the QoE gain of URLLC. The reason lies in that as summarized in Table \ref{tab:mapping}(b), the distribution of packet size for URLLC follows a truncated lognormal distribution with the mean value of 2 MByte, which is far larger than those of VoLTE and Video services. Given the larger transmission volume and strictly lower latency requirement, it is far more difficult to satisfy the QoE of URLLC. In this case, it is still satisfactory that DQL outperforms the other competitive schemes to render higher QoE gain of URLLC at a slight cost of spectrum efficiency (SE). Meanwhile, Fig. \ref{fig:compare}(d) and Fig. \ref{fig:compare}(e) demonstrate the allocation results for the demand-prediction based schemes and show significantly inferior performance, since Fig. \ref{fig:learning_r}(a)$\sim$\ref{fig:learning_r}(c) and Fig. \ref{fig:learning_r}(g)$\sim$\ref{fig:learning_r}(i) show the number of video packets dominates the transmission and simple packet-number based prediction could not capture the complicated relationship between demand and QoE. On the other hand, Fig. \ref{fig:compare}(g) illustrates that this QoE advantage of DQL comes at the cost of a decrease in SE. Recalling the definition of the reward in DQL, if we decrease the QoE weight from 5000 to 1, DQL could learn another bandwidth allocation policy (in Fig. \ref{fig:compare}(c)) yielding a larger SE yet a lower QoE. Fig. \ref{fig:compare}(g) $\sim$ \ref{fig:compare}(j) further summarize the performance comparison in terms of SE or QoE satisfaction ratios, where the vertical errorbars show the standard derivation. These subfigures validate the DQL's flexibility and advantage in resource-limited scenarios to ensure the QoE per user.

\subsection{Priority-based Scheduling in Common VNFs}
\begin{figure*}[!t]
	\centering
	\subfigure[DQL-based Prioritied Scheduling]{
		\label{fig:priority_dql} 
		\includegraphics[width=.475\textwidth]{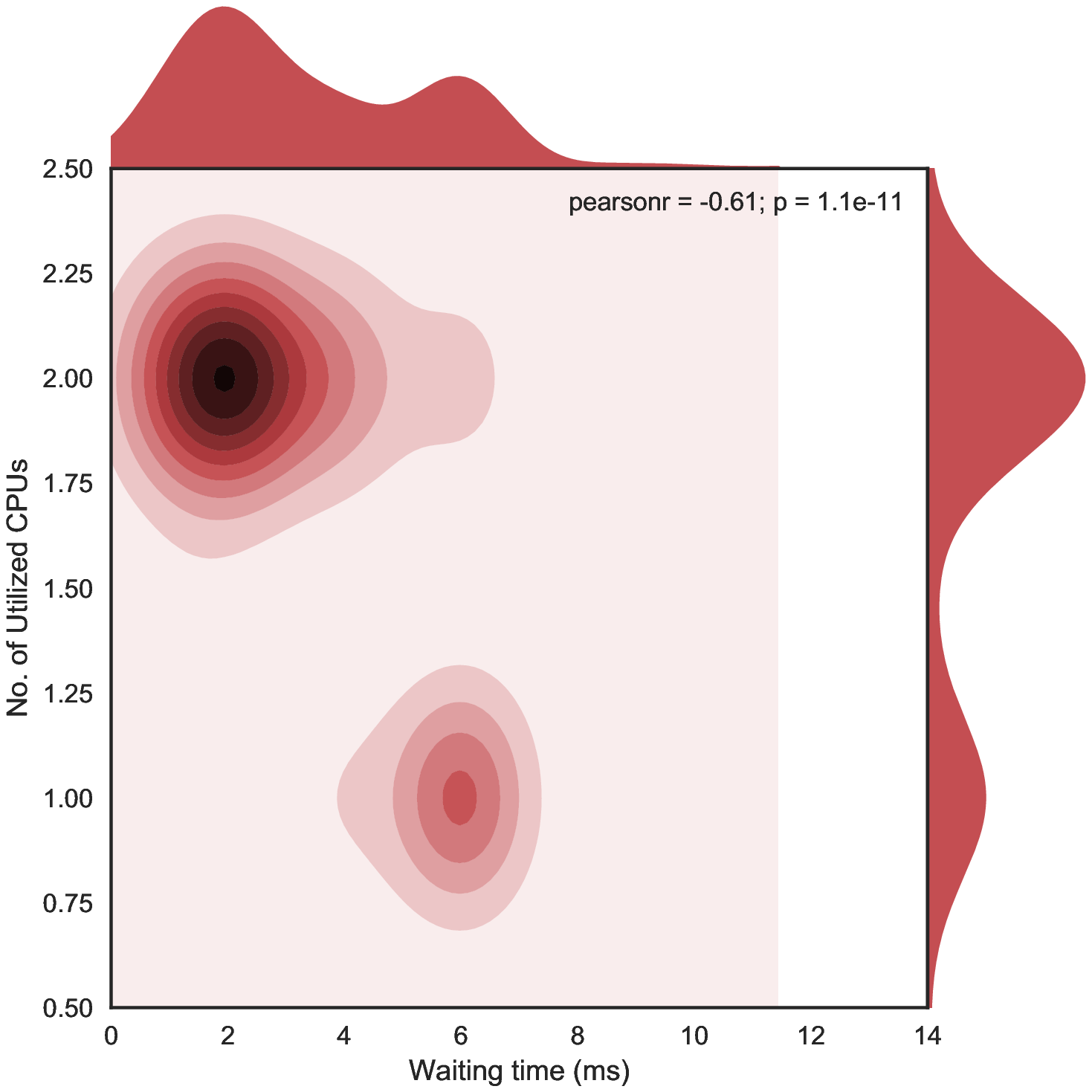}}
	\subfigure[No Priority Scheduling]{
		\label{fig:priority_every} 
		\includegraphics[width=.475\textwidth]{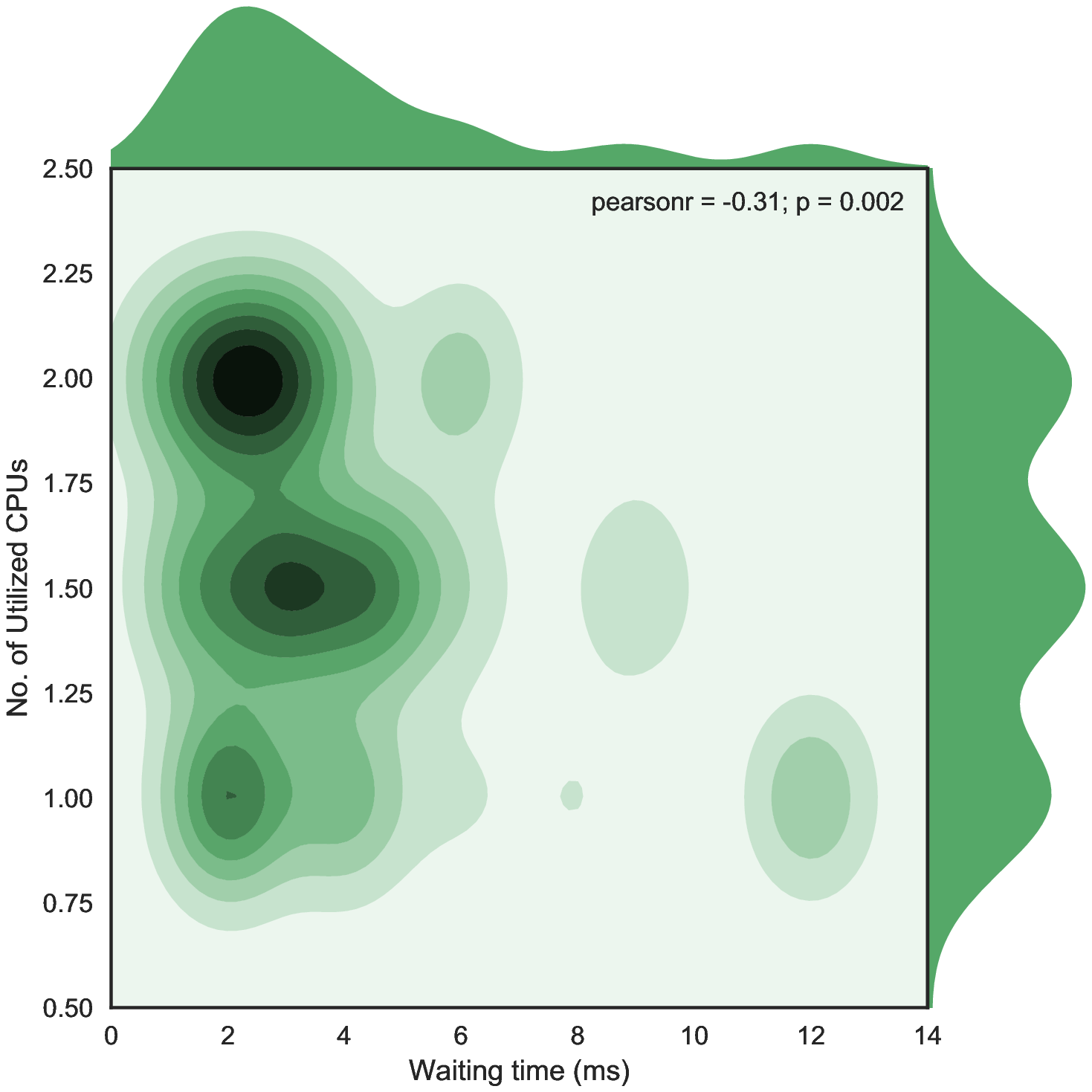}}
	\caption{Performance comparison between DQL-based priority scheduling and no priority scheduling for core network slicing.}
	\label{fig:priority} 
\end{figure*}

Section \ref{sec:radio} has discussed how to apply DRL in radio resource slicing. Similarly, if we virtualize the computation resources as VNFs for each slice, the problem to allocate computation resources to each VNF could be solved similar to the radio resource slicing case. Therefore, in this part, we talk about another important issue, that is, priority-based core network slicing for common VNFs. Specifically, we simulate a scenario where there exists 3 service function chains (SFCs) possessing the same basic capability but working at the expenditure of different computation processing units (CPUs) and yields different provisioning results (e.g., waiting time). Also, based on the commercial value or related SLA, flows could be classified into 3 categories (e.g., Category A, B, and C) with decreasing priority from Category A to Category C, and a priority-based scheduling rule is defined as that SFC I prioritizes Category A flows over the others, while SFC II equally treats Category A and B users but serves Category C flows with lower priority. SFC III treats all flows equally. Besides, SFCs process flows with equal priority according to the arrival time. The eventually utilized CPUs of each SFC depend on the number of its processed flows. Besides, SFC I, II and III cost 2, 1.5, and 1 CPU(s), but incur 10, 15, and 20 ms regardless of the flow size, respectively. Hence, subject to the limited number of CPUs, flows for each type will be scheduled to an appropriate SFC, so as to incur acceptable waiting time. Therefore, the scheduling of flows should match and learn the arrival of flows in three categories, and DQL is considered as a promising solution.

Similarly, it is critical to design an appropriate mapping of DRL elements to this slicing issue. As Table \ref{tab:mapping}(a) implies, we use a mapping slightly different from that for radio resource slicing, so as to manifest the flexibility of DQL. In particular, we abstract the state of DQL as a summary of the category and arrival time of last 5 flows and the category of the newly arrived flow, while the reward is defined as the weighted summation of processing and queue time of this flow, where a larger weight in this summation is adopted to reflect the importance of flows with higher priority. Also, we first pre-train its NN by emulating some flows with lognormal distributed inter-arrival time from the three categories' users.

We compare the DQL scheme with an intuitive ``\textit{no priority}" solution, which allocate the flow to the SFC yielding minimum waiting time. Fig. \ref{fig:priority} provides the related performance by randomly generating 10000 flows and provisioning accordingly, where the vertical and horizontal axes represent the number of utilized CPUs and the waiting time of flows respectively. Specifically, the bi-dimensional shading color reflects the number of flows corresponding to the specific waiting time and utilized CPUs. In particular, the darker color implies the larger number. Compared with the ``no priority" solution, the DQL-empowered slicing results provision flows with smaller average waiting time (i.e., 10.5\% lower than ``no priority") and significantly more sufficient CPU usage (i.e., 27.9\% larger than ``no priority"). In other words, DQL could support alternative solutions to exploit the computing resources and reduce the waiting time by first serving the users with higher commercial value. 

\section{Conclusion and Future Directions}
\label{sec:conclusion}
From the discussions in this article, we found that matching the allocated resource to slices with the users' activity demand will be the most critical challenge for effectively realizing network slicing, while DRL could be a promising solution. Starting with the introduction of fundamental concept for DQL, one typical type of DRL, we explained the working mechanism and application motivation of DQL to solve this problem. We further demonstrated the advantage of DQL in managing this demand-aware resource allocation in two typical slicing scenarios including radio resource slicing and priority-based core network slicing through extensive simulations. Our results showed that compared with the demand prediction-based and some other intuitive solutions, DQL could implicitly incorporate more deep relationship between demand (i.e., user activities) and supply (i.e., resource allocation) in resource-constrained scenarios, and enhance the effectiveness and agility for network slicing. Finally, in order to fulfill the application of DQL in a broader sense, we pointed out some noteworthy issues. We believe DRL could play a crucial role in network slicing in the future.

However, network slicing involves many aspects and a successful application of DQL needs some careful considerations: (a) Slice admission control on incoming requests for new slices: the success of network slicing implies a dynamic and agile slice management scheme. Therefore, if requests for new slices emerge, how to apply DQL is also an interesting problem since the defined state and action space requires to adapt to the changes in the ``slice" space.
(b) Abstraction of states and actions: Section \ref{sec:resource} has provided two ways to abstract state and action. Both methods sound practical in the related scenarios and reflect the flexibility of DQL. Hence, for new scenarios, it becomes an important issue to choose appropriate abstraction of states and actions, so as to better model the problem and save the learning cost. Up to date, it remains an open question on how to give some abstraction guidelines. (c) Latency and accuracy to retrieve rewards: The simulations in Section \ref{sec:resource} has assumed the instantaneous and accurate acquirement of rewards for a state-action pair. But, such an assumption no longer holds in practical complex wireless environment, since it takes time for user equipment to report the information and the network may not successfully receive the feedback. Also, similar to the case for state and action, the abstraction of reward might be difficult and the defined reward should be as simple as possible. (d) Policy learning cost: The time-varying nature of wireless channel and user activities requires a fast policy-learning scheme. However, the current cost of policy training still lacks the necessary learning speed. For example, our pre-training for the priority-based network slicing policy takes two days in an Intel Core i7-4712MQ processor to converge the Q-value function. Though GPU could speedup the training process, the learning cost is still heavy. Therefore, there are still a lot of interesting questions to be addressed.
\section*{Acknowledgment}
The authors would like to express their sincere gratitude to Chen Yu and Yuxiu Hua of Zhejiang University for the valuable discussions to implement part of simulation codes.
\bibliographystyle{IEEEtran}
\bibliography{my}

\section*{Biographies}
\begin{IEEEbiographynophoto}{Rongpeng Li} is now an assistant professor in College of Information Science and Electronic Engineering, Zhejiang University, Hangzhou China. He received his Ph.D and B.E. from Zhejiang University, Hangzhou, China and Xidian University, Xi’an, China in June 2015 and June 2010 respectively, both as “Excellent Graduates”. Dr. Li was a research engineer in Wireless Communication Laboratory, Huawei Technologies Co. Ltd., Shanghai, China from August 2015 to September 2016. He returned to academia in November 2016 as a postdoctoral researcher in College of Computer Science and Technologies, Zhejiang University, Hangzhou, China, which is sponsored by the National Postdoctoral Program for Innovative Talents. His research interests currently focus on Reinforcement Learning, Data Mining and all broad-sense network problems(e.g., resource management, security, etc) and he has authored/coauthored several papers in the related fields. He serves as an Editor of China Communications.\end{IEEEbiographynophoto}

\begin{IEEEbiographynophoto}{Zhifeng Zhao} is an Associate Professor at the Department of Information Science and Electronic Engineering, Zhejiang University, China. He received the Ph.D. degree in Communication and Information System from the PLA University of Science and Technology, Nanjing, China, in 2002. Prior to that, he received the Master degree of Communication and Information System in 1999 and Bachelor degree of Computer Science in 1996, from the PLA University of Science and Technology, respectively. From September 2002 to December 2004, he acted as a postdoctoral researcher at the Zhejiang University, where his researches were focused on multimedia NGN (next-generation networks) and soft-switch technology for energy efficiency. From January 2005 to August 2006, he acted as a senior researcher at the PLA University of Science and Technology, Nanjing, China, where he performed research and development on advanced energy-efficient wireless router, Ad Hoc network simulator and cognitive mesh networking test-bed. His research area includes cognitive radio, wireless multi-hop networks (Ad Hoc, Mesh, WSN, etc.), wireless multimedia network and Green Communications. 
\end{IEEEbiographynophoto}

\begin{IEEEbiographynophoto}{Qi Sun} received her Ph.D. degree in information and communication engineering from Beijing University of Posts and Telecommunications in 2014. After graduation, she joined the Green Communication Research Center of the China Mobile Research Institute. Her research interest focuses on 5G communications, including new waveforms, non-orthogonal multiple access, massive MIMO, full duplex.
\end{IEEEbiographynophoto}

\begin{IEEEbiographynophoto}{Chih-Lin I} is CMCC Chief Scientist of Wireless Technologies, launched 5G R\&D in 2011, and leads C-RAN, Green and Soft initiatives. Chih-Lin received IEEE Trans. COM Stephen Rice Best Paper Award, and IEEE ComSoc Industrial Innovation Award. She was on IEEE ComSoc Board, GreenTouch Executive Board, WWRF Steering Board, M\&C Board Chair, and WCNC SC Founding Chair. She is on IEEE ComSoc SPC and EDB, ETSI/NFV NOC, and Singapore NRF SAB.
\end{IEEEbiographynophoto}

\begin{IEEEbiographynophoto}{Chenyang Yang} received her Ph.D. degrees in Electrical Engineering from Beihang University (formerly Beijing University of Aeronautics and Astronautics, BUAA), China, in 1997. She has been a full professor with the School of Electronics and Information Engineering, BUAA since 1999. She has published over 200 papers and filed over 80 patents in the fields of energy efficient transmission, URLLC, wireless local caching, CoMP, interference management, cognitive radio, and relay, etc. She was supported by the 1st Teaching and Research Award Program for Outstanding Young Teachers of Higher Education Institutions by Ministry of Education of China. She was the chair of Beijing chapter of IEEE Communications Society during 2008-2012, and the MDC chair of APB of IEEE Communications Society during 2011-2013. She has served as TPC Member, TPC co-chair or Track co-chair for IEEE conferences. She has ever served as an associate editor for IEEE Trans. on Wireless Communication, guest editor for IEEE Journal of Selected Topics in Signal Processing and IEEE Journal of Selected Areas in Communications. Her recent research interests lie in mobile AI, wireless caching, and URLLC.
\end{IEEEbiographynophoto}

\begin{IEEEbiographynophoto}{Xianfu Chen}  received his Ph.D. degree in Signal and Information Processing, from the Department of Information Science and Electronic Engineering at Zhejiang University, Hangzhou, China, in March 2012. He is currently a Senior Scientist with the VTT Technical Research Centre of Finland Ltd, Oulu, Finland. His research interests cover various aspects of wireless communications and networking, with emphasis on network virtualization, software-deﬁned radio access networks, green communications, centralized and decentralized resource allocation, and the application of machine learning to cognitive radio networks. He is an IEEE member.
\end{IEEEbiographynophoto}

\begin{IEEEbiographynophoto}{Minjian Zhao} received the M.Sc. and Ph.D. degrees in communication and information systems from Zhejiang University, Hangzhou, China, in 2000 and 2003, respectively. He is currently a Professor with the College of Information Science and Electronic Engineering, Zhejiang University. His research interests include modulation theory, channel estimation and equalization, and signal processing for wireless communications.
\end{IEEEbiographynophoto}

\begin{IEEEbiographynophoto}{Honggang Zhang} is currently a Full Professor with the College of Information Science and Electronic Engineering, Zhejiang University, Hangzhou, China. He was an Honorary Visiting Professor with the University of York, U.K and an International Chair Professor of Excellence for Universit\'e Europ\'eenne de Bretagne and Sup\`elec, France. He was the Co-Author and an Editor of two books ``Cognitive Communications-Distributed Artificial Intelligence (DAI), Regulatory Policy and Economics, Implementation (John Wiley \& Sons)" and ``Green Communications: Theoretical Fundamentals, Algorithms and Applications (CRC Press)", respectively. He is also active in the research on green communications and was the leading Guest Editor of the IEEE Communications Magazine special issues on Green Communications. He is taking the role of an Associate Editor-in-Chief of China Communications and the Series Editors of the IEEE Communications Magazine for its Green Communications and Computing Networks Series. He served as the Chair of the Technical Committee on Cognitive Networks of the IEEE Communications Society from 2011 to 2012.	
\end{IEEEbiographynophoto}

\end{document}